\documentclass{llncs}
\usepackage{float}
\usepackage{tabularx,ragged2e}
\usepackage{amsmath}
\usepackage{booktabs}
\usepackage{multirow}
\usepackage{subfigure}
\usepackage{lscape}
\usepackage[table,xcdraw]{xcolor}
\usepackage{makeidx}  
 \usepackage{multirow}
 \usepackage[table,xcdraw]{xcolor}
\usepackage{amssymb}
\usepackage{array}
\newcolumntype{P}[1]{>{\centering\arraybackslash}p{#1}}
\usepackage{graphicx}
\usepackage{tabu}
\usepackage[flushleft]{threeparttable}
\usepackage{url}
\makeindex
\begin{document}

\title{Development of Diabetic Foot Ulcer Datasets: An Overview}

\author{Moi Hoon Yap\inst{1}\orcidID{0000-0001-7681-4287} 
\and Connah Kendrick \inst{1}\orcidID{0000-0002-3623-6598} 
\and Neil D. Reeves \inst{4}\orcidID{0000-0001-9213-4580}
\and Manu Goyal \inst{3}\orcidID{0000-0002-9201-1385} 
\and Joseph M. Pappachan \inst{2}\orcidID{0000-0003-0886-5255} 
\and Bill Cassidy\inst{1}\orcidID{0000-0003-3741-8120} }
%
\authorrunning{M.H. Yap et al.}
%

\institute{Centre for Advanced Computational Science, Department of Computing and Mathematics, Manchester Metropolitan University, Manchester M1 5GD, United Kingdom  \and
Lancashire Teaching Hospitals NHS Trust, 
Preston, PR2 9HT, United Kingdom \and
Department of Radiology, UT Southwestern Medical Center, 5323 Harry Hines Blvd., Dallas, Texas 75390-9085, USA \and
Musculoskeletal Science and Sports Medicine Research Centre, Manchester Metropolitan University, Manchester M1 5GD, United Kingdom  \\
\email{M.Yap@mmu.ac.uk}
}

\maketitle


\begin{abstract}
This paper provides conceptual foundation and procedures used in the development of diabetic foot ulcer datasets over the past decade, with a timeline to demonstrate progress. We conduct a survey on data capturing methods for foot photographs, an overview of research in developing private and public datasets, the related computer vision tasks (detection, segmentation and classification), the diabetic foot ulcer challenges and the future direction of the development of the datasets. We report the distribution of dataset users by country and year. Our aim is to share the technical challenges that we encountered together with good practices in dataset development, and provide motivation for other researchers to participate in data sharing in this domain.
\end{abstract}

\section{Introduction}
The Diabetic Foot Ulcer (DFU) is one of the major complications resulting from diabetes, which can lead to lower limb amputation \cite{armstrong1998validation}. Regular foot check by clinical professionals is required for patients with DFU development, which is often costly and / or requires referral to specialist care \cite{prompers2008delivery}. Research shows that healthcare services that treat DFU are unable to handle the growing number of patients due to inadequately trained medical staff \cite{cavanagh2012cost}, which is especially prevalent in low-income countries and rural areas \cite{zimmet2014diabetes,vinicor1998public}.

\begin{figure}[!htb]
\centering
\includegraphics[width=1.0\textwidth]{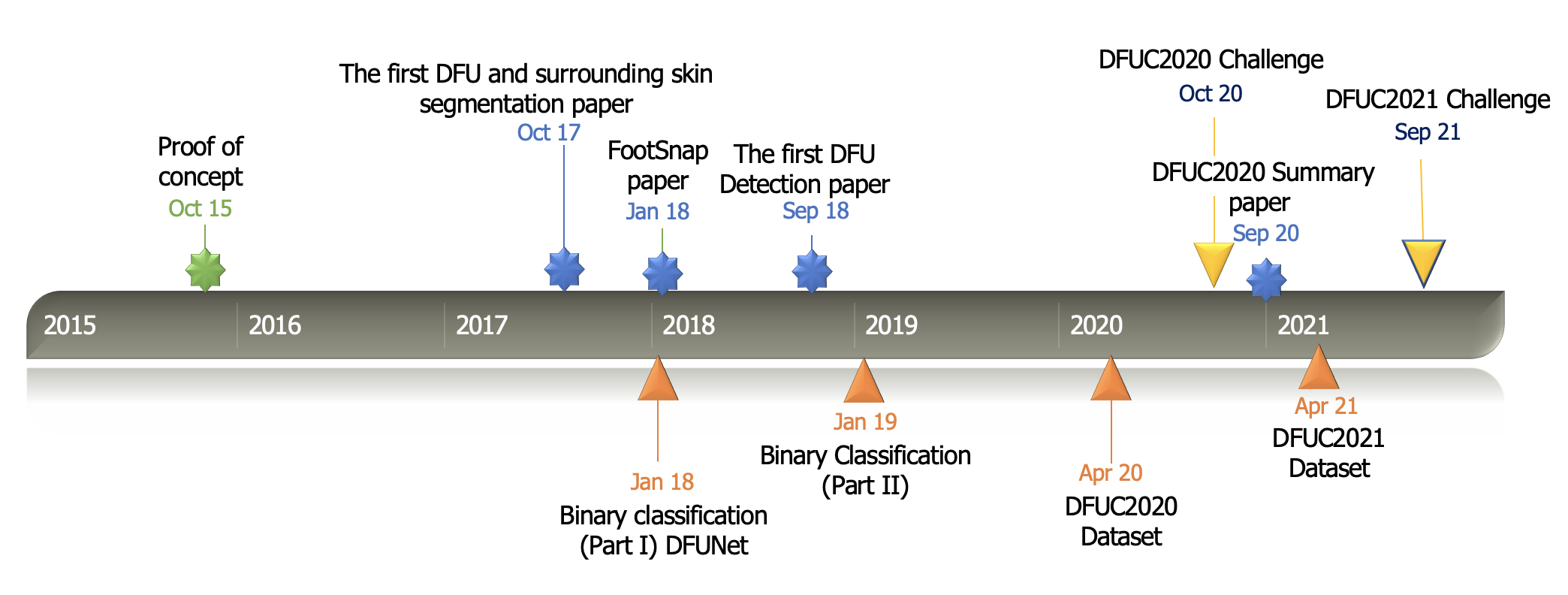}
\caption{The timeline of the development of DFU datasets and analysis. The proof-of-concept in using computer vision for DFU analysis was published in October 2015 by Yap et al. \cite{yap2015computer}. Although the FootSnap app was created in 2015, the paper was not published until 2018 \cite{yap2018new}.}
\label{figure: Timeline}
\end{figure}

Over the past decade, the development of digital and information technology has enabled the creation of new computer-based solutions for healthcare, including wound care \cite{chanussot2013telemedicine}. Figure \ref{figure: Timeline} illustrates the timeline of development of DFU datasets, including the first use of computer vision methods in DFU detection. The focus of the earlier DFU research is in the design of new capturing tools \cite{yap2018new}, initiated in 2015. At the same time, computer vision methods (based on image processing algorithms) and conventional machine learning methods were used to analyse those images \cite{yap2015computer}. With the advent of deep learning in computer vision tasks, researchers began investigating the use of deep learning for DFU segmentation in 2016, where the first fully automated segmentation paper was published in 2017 by Goyal et al. \cite{goyal2017fully}. 

Motivated by the success of deep learning on DFU segmentation, the team at the Manchester Metropolitan University and Lancashire Teaching Hospitals NHS Foundation Trust had obtained ethical approval from the UK National Health Service Research Ethics Committee (reference number: 15/NW/0539) to create larger scale datasets of DFU images, with approval to share the datasets with the research community for research, provided that the users abide by the licence agreement. The first dataset (Part I) was on binary classification using normal and DFU patches, where the authors designed a new deep learning network (DFUNet) and benchmarked the datasets with popular networks at that time \cite{goyal2018dfunet}. The second dataset (Part II) was also on binary classification, and focussed on ischaemic and infection skin patches. The binary classification of ischaemia-vs-all and infection-vs-all were benchmarked by Goyal et al. \cite{goyal2020recognition}, and they proposed an ensemble method to increase the accuracy of infection and ischaemia recognition. In 2020, the team conducted the first inaugural research challenge in DFU detection, DFUC2020 \cite{yap2020zenodo}, and the second challenge in DFU multi-class classification in 2021 \cite{yap2021zenodo}.

The remaining sections of the paper are organised as follow: Section 2 provides a survey of DFU data capturing methods; Section 3 reviews the available DFU datasets; Section 4 describes the DFU research challenges conducted over the past years; Section 5 presents future work and research directions of DFU analysis; and Section 6 summarises the paper.

\section{A Survey of Data Capturing Methods}
In current clinical practices, podiatrists and consultants use a range of digital single-lens reflex (SLR) camera models to collect DFU photographs \cite{goyal2018dfunet,goyal2018region,cassidy2020dfuc}. The photographs are transferred to a secured storage which is often isolated from the patients' electronic health records. The process is operator dependent, with poor consistency across different clinic and care settings. Due to these inconsistencies and limitations of 2D images, it has not been possible to quantify the changes of the ulcers over time.

\begin{figure}
	\centering
	\begin{tabular}{cc}
		\includegraphics[width=.4\textwidth]{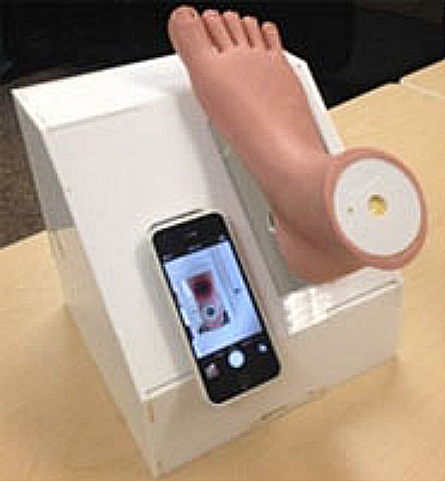} &
		\includegraphics[width=.62\textwidth]{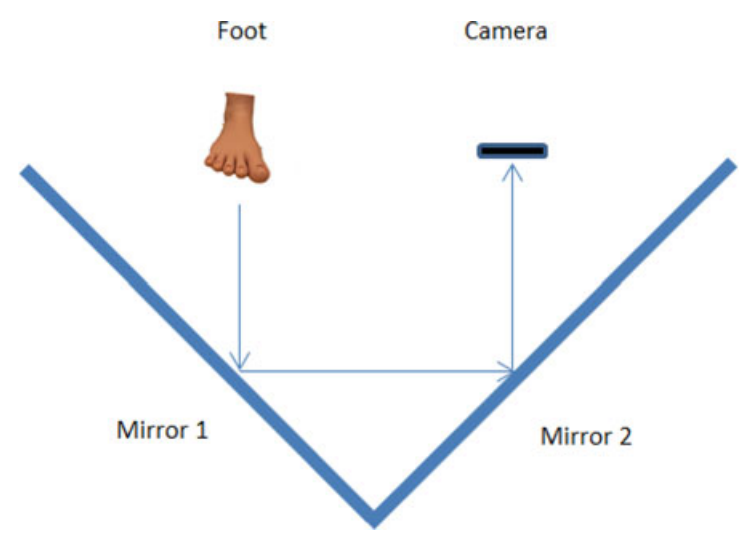}\\
		(a) & (b)\\  
	\end{tabular}
	\caption[]{Illustration of the DFU image capture box proposed by Wang et al. \cite{wang2015wound}: (a) image capture box with smartphone and foot with wound mould attached, (b) capture box optical principle.}
	\label{figure:capture_box}
\end{figure}

Over the past few years, several research teams have proposed new methods in standardising data capture of DFUs. The earliest attempts were conducted by Wang et al. \cite{wang2015wound} and Yap et al. \cite{yap2015computer}. Wang et al. \cite{wang2015wound} developed a smartphone app capable of image segmentation of DFU wounds using an accelerated mean-shift algorithm, which pre-dates current deep learning approaches. To improve and standardise the acquisition of DFU images, they created an image capture box, to be used in conjunction with the smartphone app. Their goal was to promote a more active role in patient self-monitoring. This approach used skin colour to determine foot boundaries, while wound area is determined by a simple connected region detection method. This system also assessed healing status using a red–yellow–black colour evaluation model and a quantitative trend analysis of time records for a given patient. The system was tested with 34 patients and 30 wound moulds. Figure \ref{figure:capture_box} shows the design of the capture box\footnote{reproduced with permission from Peder C. Pedersen, Dept. of Electrical and Computer Engineering, Worcester Polytechnic Institute, Worcester, MA 01609, USA}.

Wang et al. \cite{wang2015dfu} later validated the capture box in a clinical study using 32 DFU photographs obtained from 12 diabetic patients. This system used a laptop to perform wound segmentation, calculation of the wound area and calculation of a healing score. They found a good correlation between human and automated wound area measurements, reporting a Matthews correlation coefficient value of 0.68 for the wound area determination algorithm, and a Krippendorff alpha coefficient within the range of 0.42 to 0.81.

Wang et al. \cite{wang2017area} conducted a third study using their capture box, which collected 100 DFU photographs from 15 patients during a 2-year period. In this study, they utilised superpixel segmentation, using the Simple Linear Iterative Clustering (SLIC) algorithm, as inputs for a cascaded two-stage SVM-based (Support-Vector Machine) classifier to determine wound boundaries for DFU. Colour and bag-of-word representations of local dense scale invariant transformation features are used as descriptors for excluding non-wound regions. Wavelet-based features are used as descriptors to identify healthy tissue from wound tissue. Finally, wound boundaries are refined by applying a conditional random field method. This system ran on a Nexus 5 smartphone, and reported an average sensitivity of 73.3\%, and an average specificity of 94.6\% with a computation time of 15 to 20 seconds. 

\begin{figure}
	\centering
	\begin{tabular}{cc}
		\includegraphics[width=.4\textwidth]{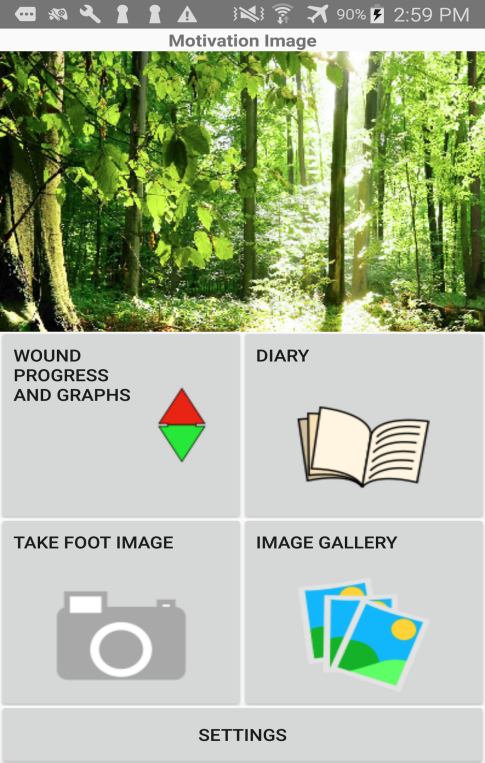} &
		\includegraphics[width=.5192\textwidth]{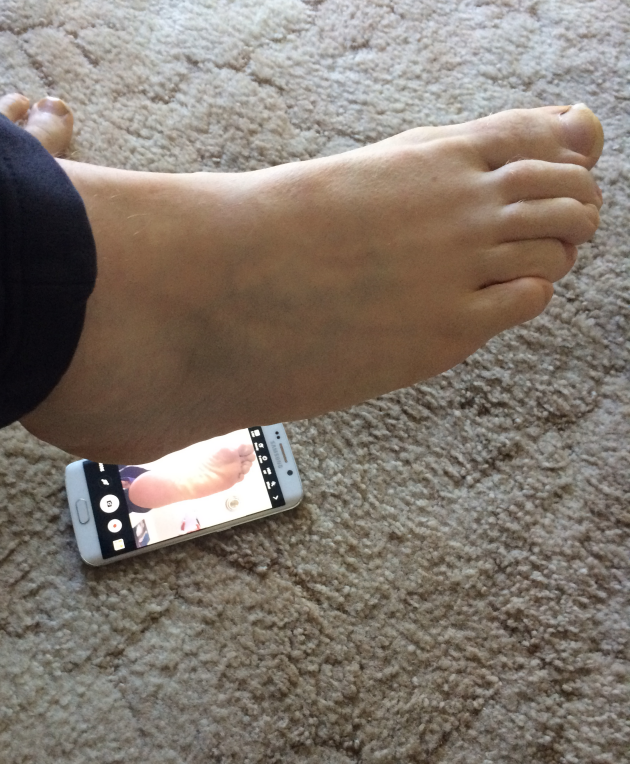}\\
		(a) & (b)\\  
	\end{tabular}
	\caption[]{Illustration of the MyFootCare smartphone app, created by Brown et al. \cite{brown2017myfootcare}: (a) main screen showing all available features, (b) demonstration of the voice-guided photograph acquisition feature.}
	\label{figure:myfootcare}
\end{figure}

In 2015, Yap et al. \cite{yap2015computer} created a new mobile app called FootSnap which was used for standardising photographic capture of the plantar aspect of feet. This system generated an outline of a foot from an initial photograph of the patient's foot. The outline could then be recalled on-screen to help align the foot in subsequent photographs. The app was initially tested on healthy feet, then later evaluated for standardisation on clinical data \cite{yap2018new}. The app was then used as part of a clinical trial investigating DFU prevention using smart insole technology published in 2019 \cite{abbott2019innovative}.

\begin{figure}
\centering
\includegraphics[width=1.0\textwidth]{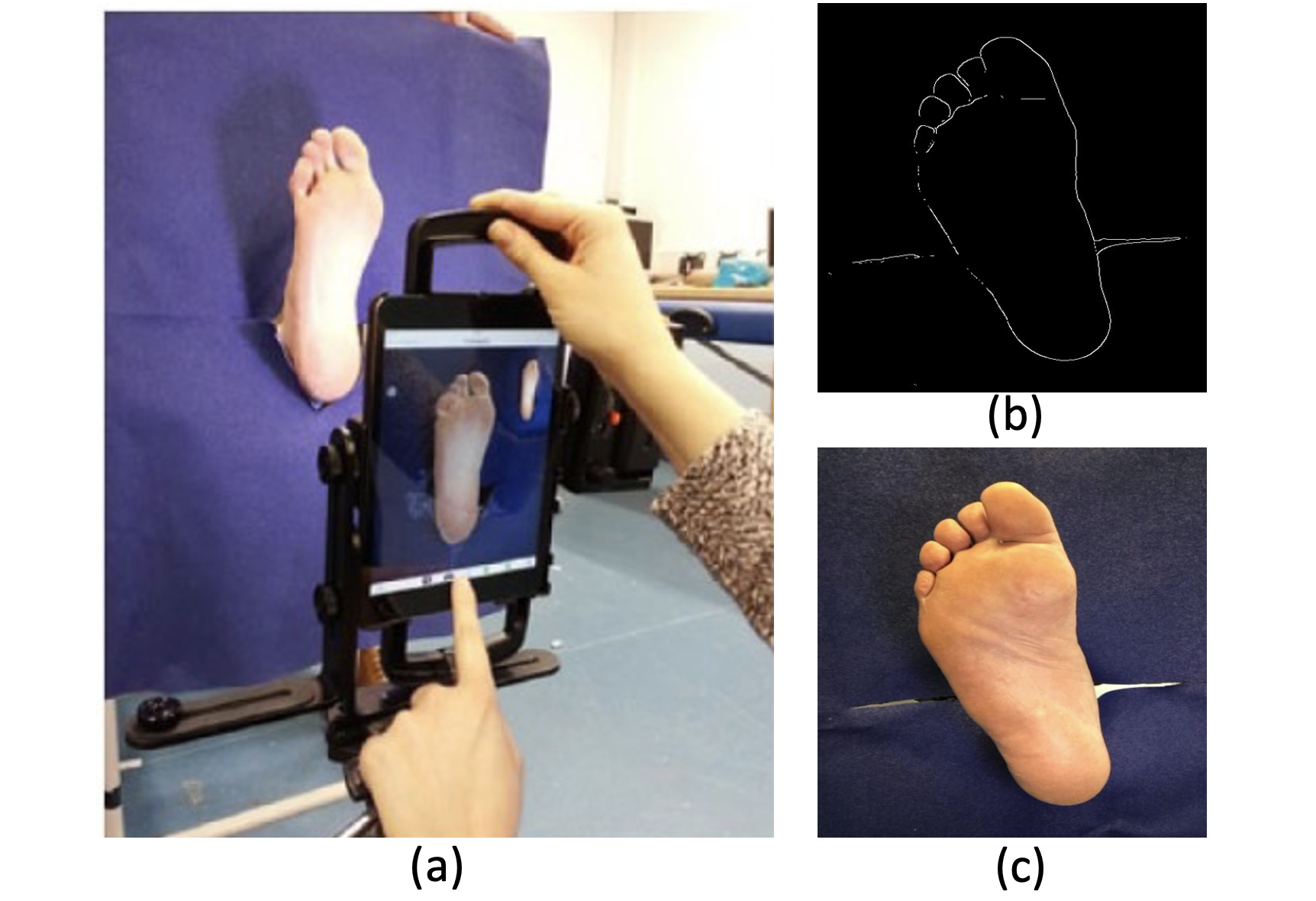}
\caption{Illustration of the FootSnap mobile app: (a) the data capturing tool and settings; (b) ghost image generated automatically to standardise data capturing; and (c) plantar aspect of foot captured.}
\label{figure: footsnap}
\end{figure}

Researchers would continue to investigate the use of mobile technologies to address the growing problem of DFU. Brown et al. \cite{brown2017myfootcare} developed the MyFootCare app - a self-monitoring tool that could be used by patients to promote self-care in home settings. The app was evaluated with 3 DFU patients, who reported it as useful for tracking wound progress and to assist in communications with clinicians. The app implements a number of features, including image capture, wound analysis (using OpenCV), diary reminders, and wound size tracking using a graph. The image capture allows the patient to place the phone onto the floor with the screen facing upwards. The patient can then guide their foot into the correct position, with voice feedback provided by the app. This allows the patient to remain seated while positioning their foot for image capture. However, this functionality was not completed in time for the evaluation with patients. Figure \ref{figure:myfootcare} shows the design of the app and its proposed use in DFU photograph acquisition\footnote{reproduced with permission from Ross Brown, School of Computer Science, Science and Engineering Faculty, Queensland University of Technology, Brisbane, Queensland, Australia}.

\begin{figure}
	\centering
	\begin{tabular}{ccc}
		\includegraphics[width=.38\textwidth]{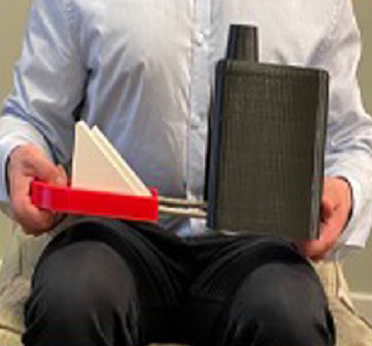} &
		\includegraphics[width=.2724\textwidth]{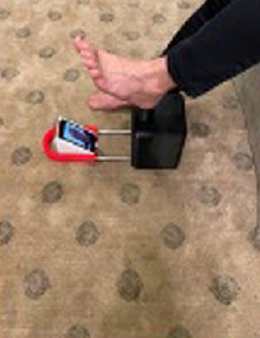} &
		\includegraphics[width=.2662\textwidth]{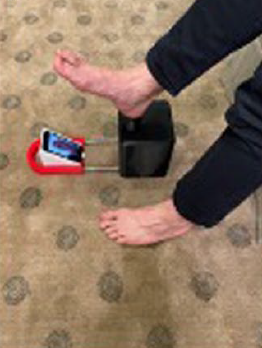} \\
		(a) & (b) & (c) \\
	\end{tabular}
	\caption[]{Illustration of the "foot selfie" DFU monitoring system proposed by Swerdlow et al. \cite{swerdlow2021selfie}: (a) side view of apparatus showing smartphone holder on the left and foot holder on the right, (b) left foot photograph acquisition, (c) right foot photograph acquisition.}
	\label{figure:selfie_box}
\end{figure}

More recent techniques for photographic acquisition of DFU have been proposed by Swerdlow et al. \cite{swerdlow2021selfie}. They devised a "foot selfie" device comprising an elaborate assembly which helps to position the foot in front of a smartphone while minimising surface contact (see Figure \ref{figure:selfie_box})\footnote{reproduced with permission from David G. Armstrong, Southwestern Academic Limb Salvage Alliance (SALSA), Department of Surgery, Keck School of Medicine of University of Southern California, Los Angeles, California, USA}. Compared to some of the earlier capture methods, this solution has the advantage of not requiring contact between wound and surface, and also allows for capture of more than just plantar DFU.

\begin{figure}
\centering
\includegraphics[width=1.0\textwidth]{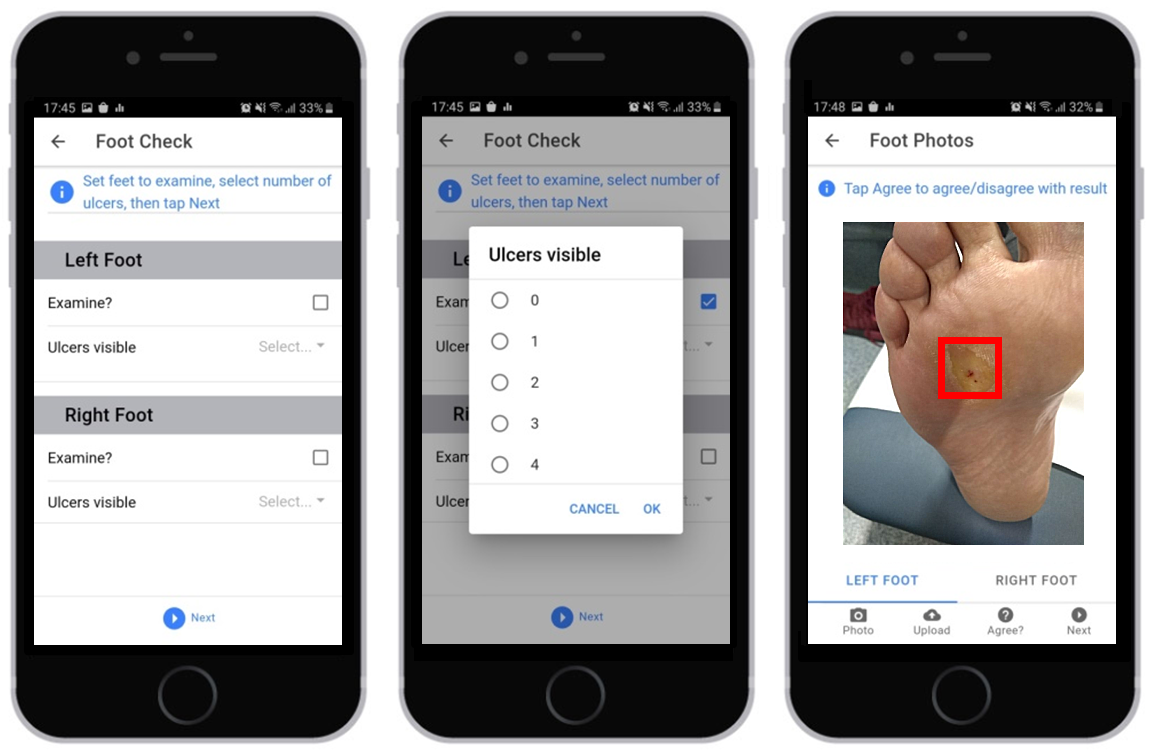}
\caption{Illustration of the cross-platform mobile app used to capture DFU photographs at two UK NHS hospitals during a six-month proof-of-concept clinical evaluation.}
\label{figure:footsnap_poc}
\end{figure}

Cassidy et al. recently validated a fully automated DFU detection system which utilised mobile and cloud-based technologies \cite{cassidy2021cloudbased,reeves2021diabetes}. This system used a cross-platform mobile app to capture photographs of patient's feet in clinical settings that could be uploaded to a cloud platform for inference to detect the presence of DFU. The system was validated in a proof-of-concept study at two UK hospitals over a six-month period. This technology is currently being adapted for use in additional clinical studies with the aim of replacing SLR cameras in the acquisition of DFU photographs. Figure \ref{figure:footsnap_poc} shows a selection of screens from the cross-platform mobile app.

\section{A Review of DFU Image Datasets}
This section reviews the available public DFU image datasets and the number of users (by countries, if known). Table \ref{tab:papers} summarises and compares five publicly available datasets.

\begin{table}
    \centering
    \renewcommand{\arraystretch}{1.0}
    \caption{A summary of DFU image datasets.}
    \label{tab:papers}
    \scalebox{1.0}{
    \begin{tabular}{|p{2.6cm}|p{0.8cm}|p{2cm}|p{1.5cm}|p{2cm}|p{1cm}|p{1cm}|p{1cm}|}
		\hline
		Publication & Year & Dataset Name & Resolution& Task & Train& Test& Total\\\hline
		Goyal et al. \cite{goyal2018dfunet} & 2018 & Part A or \newline Part I & varied & classification& NA & NA & 1,679\\ \hline
		Goyal et al. \cite{goyal2020recognition} & 2019 & Part B or \newline Part II & $256\times256$& classification & NA & NA & 1,459 \\ \hline
		Cassidy et al. \cite{cassidy2020dfuc} & 2020 & DFUC2020 & $640\times480$ & detection & 2,000 & 2,000 & 4,000 \\ \hline
		Wang et al. \cite{wang2020} & 2020 & AZH wound care dataset & $224\times224$& segmentation & 831 & 278 &1,109 \\ \hline 
		Thomas \cite{Thomas2020} & NA & Medetec & $560\times391$ \newline $224\times224$& segmentation & 152 & 8 & 160\\ \hline 
		Wang et al. \cite{wang2021zenodo} & 2021 & FUSeg Challenge &$512\times512$& segmentation & 1,010 & 200 & 1,210 \\ \hline

		Yap et al. \cite{yap2021analysis} & 2021 & DFUC2021 &$224\times224$& classification& 5,955 & 5,734 & 15,683*\\ \hline
		\multicolumn{4}{@{}l}{* 3.994 patches are unlabelled; NA indicates unknown}
		\end{tabular}
		}
	\end{table}

\subsection{Binary Classification}
\subsubsection*{DFU patches and normal patches (Part A or Part I)}
The Part A DFU Dataset \cite{goyal2018dfunet} consists of 1038 DFU patches and 641 normal patches. This is the first binary classification dataset shared with the research community. The ground truth was produced by two healthcare professionals, specialising in DFU, using the annotation tool developed by Hewitt et al. \cite{Hewitt2016}. The authors introduced this dataset and created a new deep learning model, DFUNet, to benchmark the performance on the dataset. DFUNet achieved an F1-score of 0.939. Since its initial release, the dataset has been used in later research, with the most recent publication achieving the best performance of 0.952 in F1-score \cite{nora2022diabetic}.

\subsubsection*{Recognition of Infection and Ischaemia Datasets (Part B or Part II)}
The first infection and ischaemia datasets were created by Goyal et al. \cite{goyal2020recognition}, which consists of 1,459 DFUs: 645 with infection, 24 with ischaemia, 186 with infection and ischaemia, and 604 control DFU (presence of DFU, but without infection or ischaemia). This work focused on binary classification, i.e., infection-vs-all and ischaemia-vs-all. This dataset consists of:

\begin{itemize}
    \item 4,935 patches (including augmented images) of ischaemia
    \item 4,935 patches (including augmented images) of non-ischaemia
    \item 2,946 patches (including augmented images) of infection
    \item 2,946 patches (including augmented images) of non-infection
\end{itemize}
\noindent
Image labelling information: 
\begin{itemize}
    \item 00XXXX\_1X.jpg  - the original image patch
    \item 00XXXX\_2X,jpg and 00XXX\_3X.jpg - natural data augmentation, where M indicates mirroring; R1, R2 and R4 indicate rotations.
\end{itemize}

\begin{figure}
	\centering
	\begin{tabular}{cccc}
		\includegraphics[width=.25\textwidth]{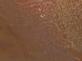} &
		\includegraphics[width=.25\textwidth]{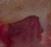} &
		\includegraphics[width=.25\textwidth]{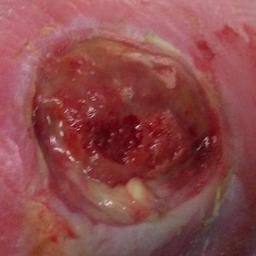}&
		\includegraphics[width=.25\textwidth]{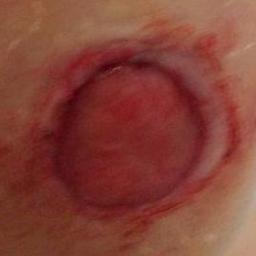} \\
		(a) & (b) & (c) &(d)\\
	\end{tabular}
	\caption[]{Comparison of image patches from Part A and Part B datasets: (a) normal patch from Part A, (b) abnormal patch from Part A, (c) infection patch from Part B, and (d) other patch from Part B.}
	\label{figure:Part_samples}
\end{figure}

Goyal et al. \cite{goyal2020recognition} introduced an ensemble CNN method for binary classification and achieved an F1-score of 0.902 and 0.722 on ischaemia-vs-all and infection-vs-all respectively. Since the release of this dataset in early 2020, Al-Garaawi et al. \cite{nora2022diabetic} achieved the best F1-scores of 0.990 and 0.744 on ischaemia-vs-all and infection-vs-all, respectively. This demonstrates a significant challenge for machine learning algorithms in recognising infection from other classes.

Figure \ref{figure:Part_samples} compares the Part A and Part B datasets. It is noted that the Part A dataset did not include the whole ulcer region, as it consists of image patches cropped from examples exhibiting ulcers and non-ulcers. In contrast, the Part B dataset consists of ulcers with different pathologies. 

\subsection{DFU Detection (DFUC2020)}
In 2018, Goyal et al. \cite{goyal2018robust} reported the performance of object detection algorithms on an in-house DFU dataset with 1,775 images. They proposed the use of a 2-tier transfer learning method using Faster R-CNN with InceptionV2 model. Overall, they achieved the best mean average precision (mAP) of 91.8\% on 5-fold cross-validation.

\begin{figure}[!htb]
  \centering
  \includegraphics[width=1.0\textwidth]{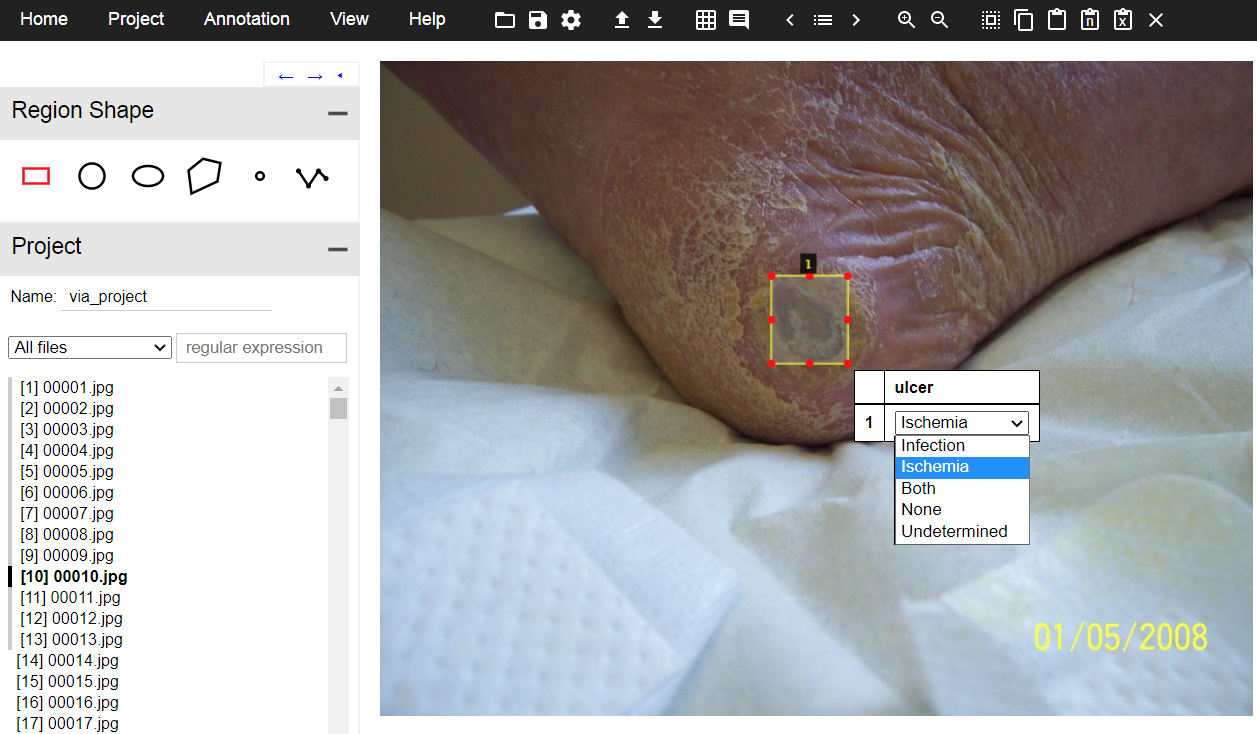}
  \caption{Illustration of expert labelling provided for the DFUC2020 dataset using the VGG Image Annotator \cite{dutta2016via}. Patches from this dataset, together with class labels would later be used in the DFUC2021 dataset.}
  \label{figure:dfuc2020_box_label}
\end{figure}

Cassidy et al. \cite{cassidy2020dfuc} introduced the largest DFU detection dataset to date, which was released on the 27th April 2020. This dataset contains 4,000 images (largely DFU images, with a small proportion of non-DFU images in the testing set), is highly heterogeneous and includes numerous challenging examples to help ensure robustness in algorithm development. Figure \ref{figure:dfuc2020_box_label} shows an example of the expert labelling provided by podiatrists for this dataset.

\subsection{Multi-class DFU Classification (DFUC2021)}
The first multi-class DFU classification dataset was released on the 27th April 2021 by Yap et al. \cite{yap2021analysis}, with four types of DFU patches, i.e., control, ischaemia, infection and both (co-occurrence of ischaemia and infection). This dataset is comprised of cropped ulcer regions from the DFUC2020 dataset \cite{cassidy2020dfuc} and the Part B dataset \cite{goyal2020recognition}.
\begin{figure}
\centering
\includegraphics[width=1.0\textwidth]{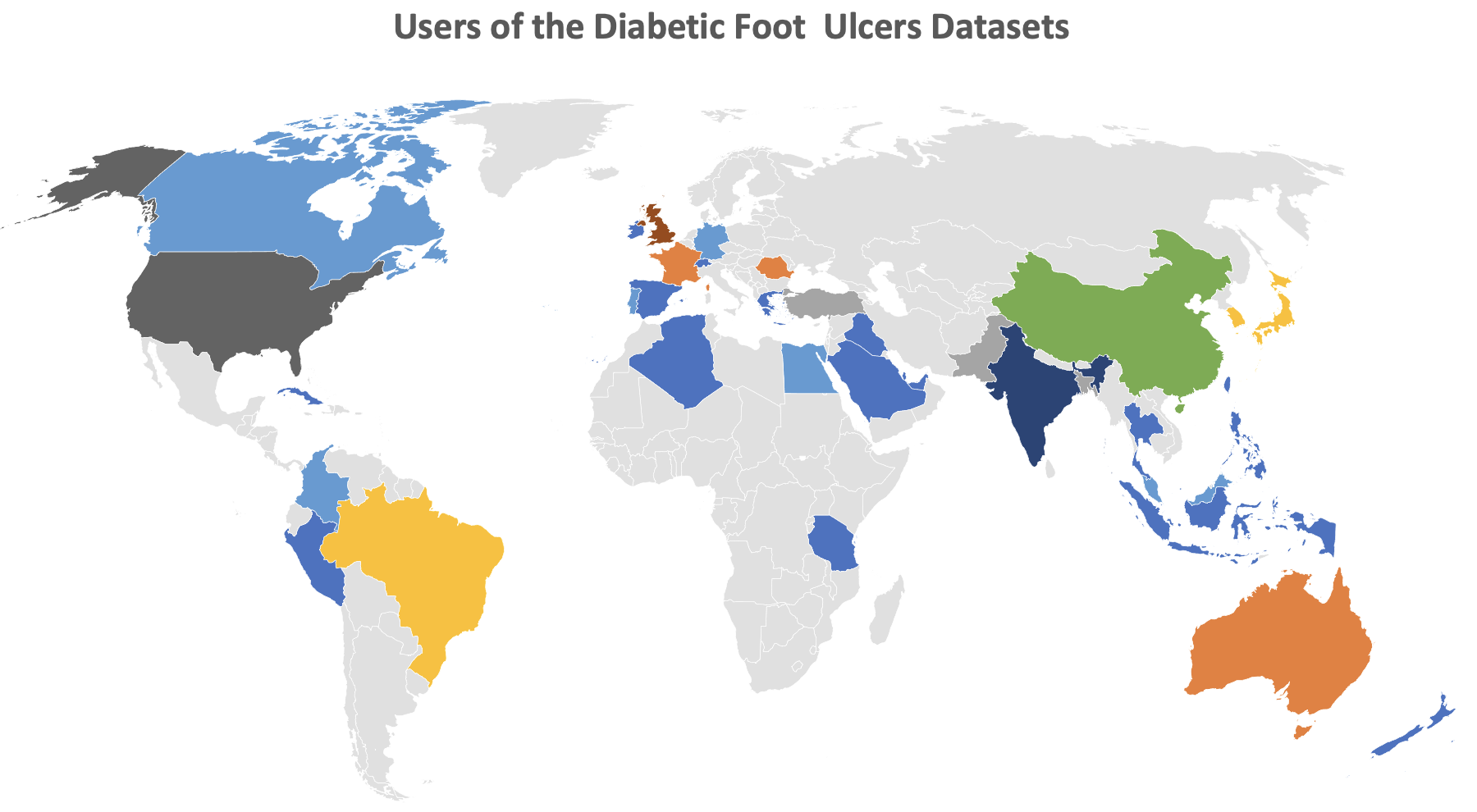}
\caption{Distribution of researchers using the DFUC2020 and DFUC2021 datasets by country. At the time of writing this paper, the datasets were requested by users from 38 countries. This map generation is powered by Bing.}
\label{figure: Datasets_countries}
\end{figure}

\begin{figure}
\centering
\includegraphics[width=1.0\textwidth]{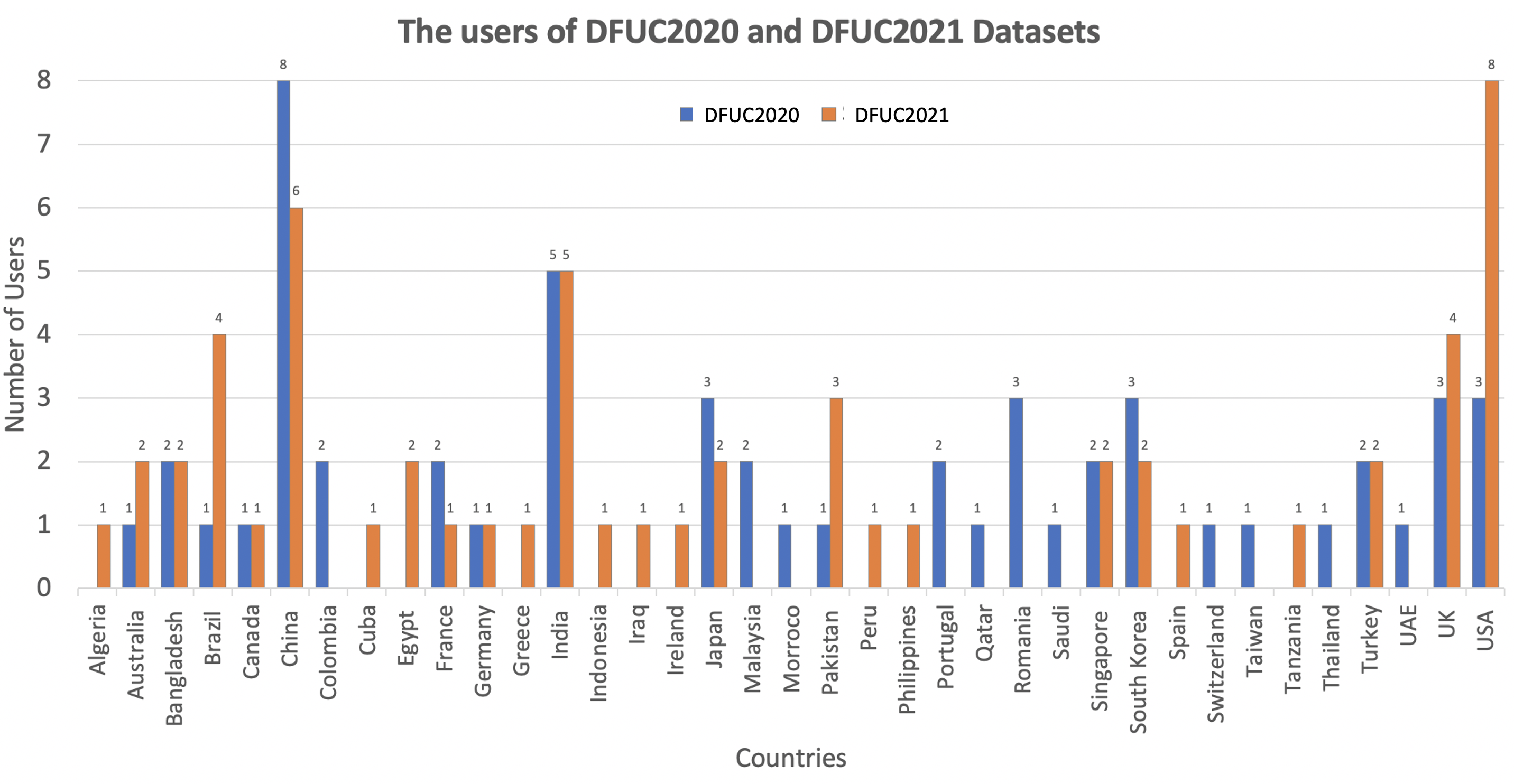}
\caption{Distribution of researchers using the DFUC2020 and DFUC2021 datasets and their country of origin. It is noted that DFUC2021 has reached wider research community, including researchers from Algeria, Cuba, Egypt, Greece, Indonesia, Iraq, Ireland, Peru, Philippines, Spain and Tanzania. Please note that we did not include New Zealand in this figure, as it was used by clinicians, rather than the challenge.}
\label{figure: Datasets_users}
\end{figure}

Since the release of the DFU Challenge datasets on the 27th April 2020, our datasets have been requested by users from 38 countries, as illustrated in Figure \ref{figure: Datasets_countries}. To date, there are more requests from China, US, India and UK. When compared with the users of DFUC2020 and DFUC2021, we observe that DFUC2021 reached a wider research community as shown in Figure \ref{figure: Datasets_users}, including growing interest of researchers from Algeria, Cuba, Egypt, Greece, Indonesia, Iraq, Ireland, Peru, Philippines, Spain and Tanzania.

\subsection{Other Datasets}
\begin{figure}
	\centering
	\begin{tabular}{ccc}
		\includegraphics[width=.31\textwidth]{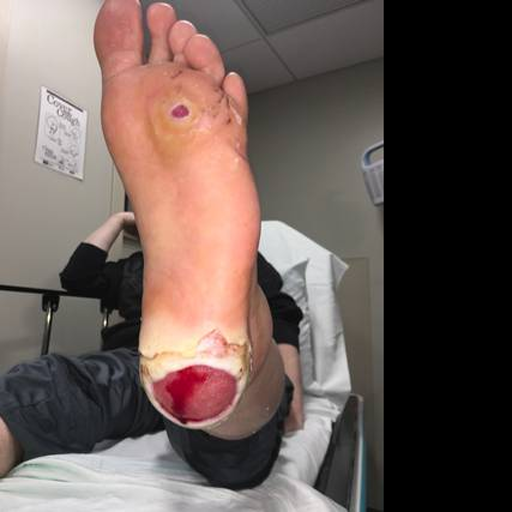} &
		\includegraphics[width=.31\textwidth]{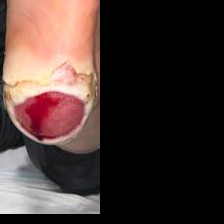} &
		\includegraphics[width=.31\textwidth]{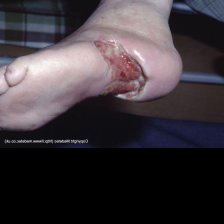} \\
		(a) & (b) & (c) \\
	\end{tabular}
	\caption[]{Illustration of the combined DFU segmentation dataset released by Wang et al. \cite{wang2020} in 2020 which comprises three sub-datasets: (a) example image from the Foot Ulcer Segmentation Challenge dataset, (b) example image from the Advancing the Zenith of Healthcare dataset, and (c) example image from the Medetec dataset. Note that the images required padding due to the non-standard size of the source images.}
	\label{figure:WoundSegDataset}
\end{figure}

Following the successful release of the Part A, Part B and DFUC2020 datasets by researchers at Manchester Metropolitan University and Lancashire Teaching Hospitals \cite{cassidy2020dfuc,yap2020zenodo,yap2020deep}, other research groups have followed this path and released their own public DFU datasets. Wang et al. \cite{wang2020} released the Foot Ulcer Segmentation Challenge dataset which focuses on single class semantic segmentation of foot ulcer wounds and contains 1,210 images with 1,010 labels, of which 200 are used for testing. Images were captured in clinical settings at different angles, with many cases exhibiting background noise. The dataset contains images of the same ulcers at different angles at different time intervals. This dataset presents additional challenges for computer vision and deep learning algorithms as all images are padded with black pixels to maintain the aspect ratio and image size of $512\times512$ pixels, as illustrated in Figure \ref{figure:WoundSegDataset}(a). 

The AZH Wound Care Center Dataset was also introduced by Wang et al. in 2020 and contains 831 training images and 278 test images. All images in this dataset are $224\times224$ pixels and contain only DFU patches. The majority of each image in this dataset contains padding using black pixels, as shown in Figure \ref{figure:WoundSegDataset}(b). A smaller dataset was also released (Medetec), as shown in Figure \ref{figure:WoundSegDataset}(c). The release date of this dataset is unknown. This is a collection of multiple wound types, and contains 46 DFU images with no labels. The image resolution of this dataset ranges from $560\times347$ pixels to $224\times444$ pixels.

The segmentation dataset introduced by Wang et al. for the Foot UlcerSegmentation Challenge is bundled with 2 other datasets. First, is the Medetec wound dataset, resized to $224\times224$ with black padding along the bottom. The Medetec dataset has 160 wound images, 152 labels and 46 DFU examples. These images focus on the wounds, but still exhibit some background details. The second dataset is the AZH wound care center dataset, which has 1,109 images and 831 labels. Similar to Medetec, these images are $224\times224$, but padded at the bottom and sides with black borders. However, these images where pre-cropped to focus on the lesion only so do not show background on the rest of the foot.


\section{DFU Challenges}
Once a year, The Medical Image Computing and Computer Assisted Intervention (MICCAI) Society conduct medical image challenges \footnote{http://www.miccai.org/special-interest-groups/challenges/miccai-registered-challenges/} to support and lead to thoughtful research challenges. The registered MICCAI challenge is reviewed and evaluated by expert panels, criteria include the design, metrics and transparency toward higher quality challenges. Due to limited capacity and an increased number of proposals, some challenges were accepted as MICCAI endorsed events, which are online only challenges and are not associated with the conference.

The inaugural DFU challenge was initiated by Yap et al. \cite{yap2020zenodo} on the DFU detection task (DFUC2020). Lead by the Manchester Metropolitan University (UK) and Lancashire Teaching Hospitals (UK), together with other co-organisers including the University of Southern California (USA), University of Waikato (New Zealand), University of Manchester and Manchester Royal Infirmary (UK), Manipal College of Health Professions (India), Baylor College of Medicine (USA) and Waikato District Health Board (New Zealand). DFUC2020 was accepted as a MICCAI registered challenge, and was conducted in conjunction with MICCAI 2020. The DFUC2020 datasets \cite{cassidy2020dfuc} include 2,000 training images and 2,000 testing images. The summary of the challenge results were concluded by Yap et al. \cite{yap2020deep}. The organisers continue to support the research community with a live leaderboard on the Grand Challenge System \footnote{https://dfu2020.grand-challenge.org}. To date, the best result on the live leaderboard reports an mAP of 0.73.

DFUC2021 \cite{yap2021zenodo} was accepted as a MICCAI registered challenge, and was conducted in conjunction with MICCAI 2021. The focus on DFUC2021 was on classification, where the DFU patches were classified into control / none, infection, ischaemia and both conditions \cite{yap2021analysis}. The organisers continue to support the research community with a live leaderboard on the Grand Challenge System \footnote{https://dfu-2021.grand-challenge.org}. At the time of writing this paper, the best macro F1-score on the live leaderboard is 0.6307 \cite{cassidy2021diabetic}.

In the same period, another research group based in the US organised an online-only Foot Ulcer Segmentation Challenge (FUSeg) \cite{wang2021zenodo}, which was conducted as a MICCAI endorsed event. The best performance of FUSeg is a Dice score of 0.8880. Since the evaluation is not on a live leaderboard and the participants were required to send their codes (docker/container) to the organiser. It is currently unclear if the organisers are still accepting submissions.

\section{Future directions}
\begin{figure}
	\centering
	\begin{tabular}{ccc}
		\includegraphics[width=.3\textwidth]{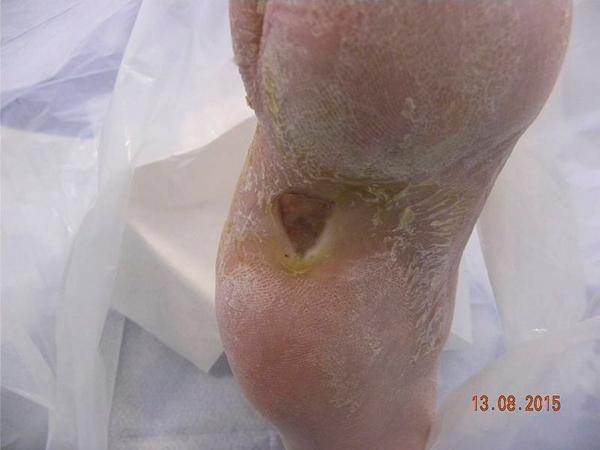} &
		\includegraphics[width=.3\textwidth]{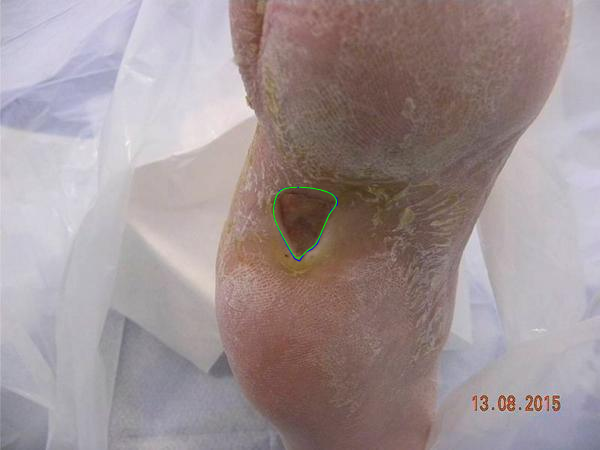} &
		\includegraphics[width=.3\textwidth]{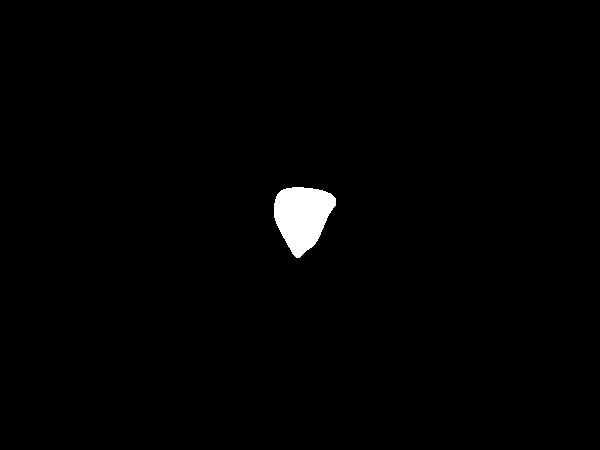}\\
		(a) & (b) &(c) \\
	\end{tabular}
	\caption[]{Illustration of an example from DFUC2022 segmentation dataset, to be released in 2022, where: (a) sample image with higher resolution, (b) delineation by the expert, and (c) the binary mask for the ulcer's region.}
	\label{figure:DFUC2022_dataset}
\end{figure}

In 2021, the DFUC2022 challenge proposal was accepted by MICCAI as a registered challenge \cite{yap2022zenodo}. The task concerns DFU segmentation \footnote{https://dfu-challenge.github.io/}. Compared to the online-only segmentation challenge conducted in 2021, DFUC2022 will comprise of large-scale higher resolution datasets, as illustrated in Figure \ref{figure:DFUC2022_dataset}. It will be conducted in conjunction with MICCAI 2022.

\begin{figure}[!htb]
\centering
\includegraphics[width=1.0\textwidth]{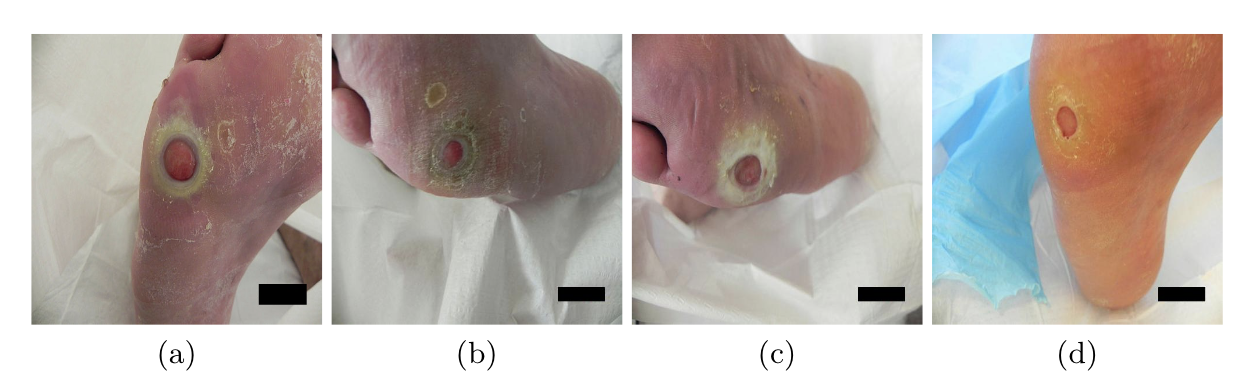}
\caption{Images acquired in a clinic at different intervals with difficulty in aligning the timeline photographs. This process is operator dependent and results in inconsistent illumination.}
\label{figure: timeline_realworld}
\end{figure}

To monitor the healing progress of DFUs in clinical settings, podiatrists / consultants take photographs of the foot using standard SLR cameras. Due to a lack of standardisation, the photographs taken are operator dependent, with variations in angle, distance and illumination. Figure \ref{figure: timeline_realworld} shows photographs of the same wound taken at different time points. 

\begin{figure}[!htb]
\centering
\includegraphics[width=1.0\textwidth]{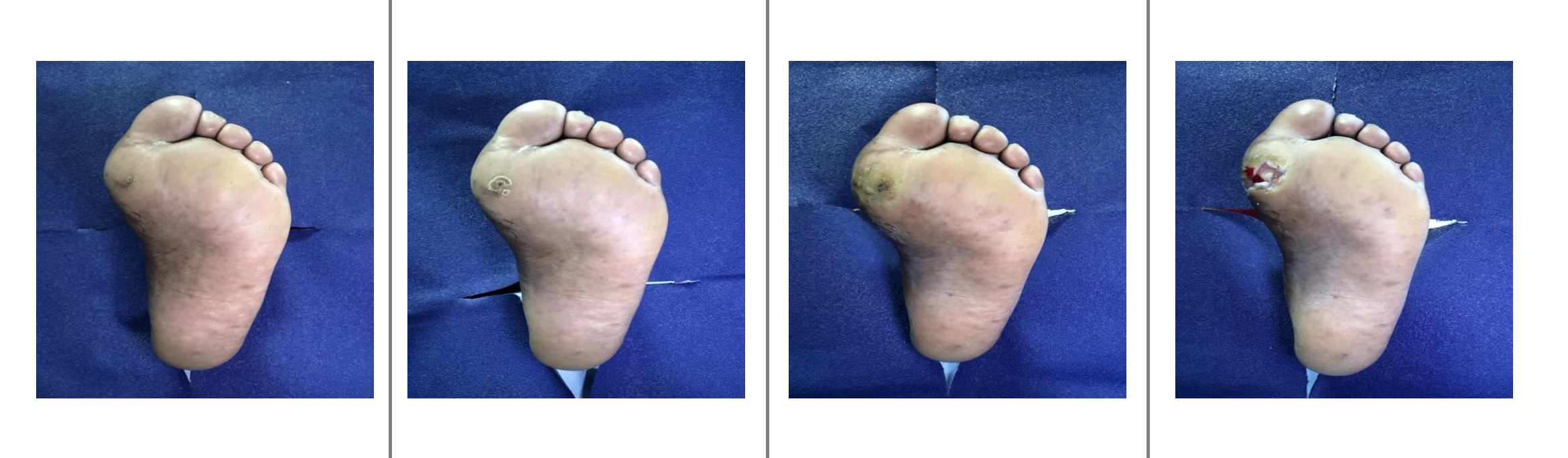}
\caption{Images taken at different intervals captured using the FootSnap mobile app, where subtle changes in the images can be observed on the forefoot.}
\label{figure: dfu_timeline}
\end{figure}

The efforts shown in previous research \cite{yap2018new} attempted to standardise the data capturing process, enabling improved observation on images captured at different timelines. Such longitudinal datasets, as illustrated in Figure \ref{figure: dfu_timeline}, help computer vision techniques and human observers to better spot the subtle changes on feet.

In an effort to improve patient care and reduce the strain on healthcare systems, current research has been focused on the development of AI algorithms that can detect diabetic foot ulcers at different stages and grades on clinical wound / DFU images. These algorithms could potentially be used as part of a mobile application that patients could use themselves (or a carer / partner) to remotely monitor their foot condition and detect the appearance of DFU for timely clinical intervention. Also, clinicians and podiatrists can monitor the progress of DFU (detected by AI algorithms) through timeline images. This is an important step in deciding what therapeutic intervention is required depending upon the progression of DFU. Effective diagnosis of such wounds can lead to better treatments, which may lead to quicker healing, reduced amputation risk and a significant reduction in health care costs. 

This paper discusses the challenges and opportunities in clinical DFU image datasets taken by different types of cameras and the role of AI algorithms in the detection and progress of DFU. Other than DFU images, researchers have used other imaging modalities such as infrared, Magnetic Resonance Imaging (MRI), and fluorescence imaging for the management of DFU. However, there are no such public datasets of other imaging modalities available for further research and development of AI solutions for multi-modality datasets.

In many research studies, thermal infrared imaging has been proven to be a useful technique in the clinical management of DFU. Several diabetic foot complications such as neuropathic ulcers, osteomyelitis and Charcot's foot have been identified at locations with increased temperature \cite{harding1998infrared,van2013infrared,harding1999infrared,armstrong2017diabetic}. Increased plantar temperature is a strong indicator of pre-ulcer conditions and may present a week before an ulcer appears visually on the foot. Hence, regular monitoring of temperature (i.e. temperature difference ($>$2.2 Celsius)) when comparing general foot temperature with suspected DFU sites can potentially help in early interventions.


Another imaging modality known as fluorescence imaging can detect the presence of clinically significant bacteria in diabetic foot ulcers by using a handheld device \cite{wu2016handheld}. Potentially, fluorescence imaging can provide valuable information of DFU outcomes on whether a DFU is healing or not \cite{lindberg2021predicting}.


MRI is the modality of choice for imaging both Charcot foot and deep infection in the diabetic foot. In the early stages, MRI can demonstrate marrow oedema while plain films remain normal. the use of MRI is common with diabetic patients to rule out infection in the presence of an ulcer, to evaluate the severity of Charcot foot, or to distinguish between Charcot foot and infection \cite{tan2007mri,schwegler2008unsuspected,forsythe2016assessment}. 

In the current literature, there are no studies that combine AI interventions for the detection and management of DFUs in multi-modality imaging. Combining wound / ulcer images with different types of other imaging (such as MRI, Thermal Infrared, and fluorescence) can potentially help AI algorithms to provide a complete diagnosis and prognosis of DFUs and timely interventions in the treatment of diabetic foot ulcers to avoid amputations. 

There are currently no publicly available datasets that combine the multi-modality imaging of diabetic foot patients. Great effort is needed for the collection of such datasets which combines different types of imaging such as thermal infrared (early detection of ulcers), clinical wound images (progression of ulcers), fluorescence (presence of clinically significant bacteria) and MRI (presence of infection and Charcot's foot). Similarly, most of the current state-of-the-art AI algorithms rely on supervised learning, hence annotation of the dataset is another important step required for the development of AI algorithms for diagnosis and management of the diabetic foot. DFU datasets are prone to the same issues that have affected other medical imaging datasets, such as image duplication, feature over-representation and sourcing of large numbers of images from a relatively small pool of subjects \cite{wen2021datasets,cassidy2021isic,daneshjou2021guidelines}.

\section{Conclusion}
This paper provides an overview of the development of DFU datasets and notable advances in the field that have led to the current use of deep learning techniques. The aim is to guide researchers in this domain to understand the breadth and depth of the processes involved in DFU classification, detection and segmentation, and to promote good practice in research and data sharing.

Collection, labelling and curation of DFU datasets is a challenging process requiring significant input from clinical experts at all stages of development. Comprehensive inter- and intra-rater analysis will prove to be key in refining the quality of datasets together with establishing new standards in this relatively new research domain. Researchers training deep learning models should pay particular attention to challenging examples within the datasets, as these will help to make networks more robust in real-world settings.

\section*{Acknowledgment}
We gratefully acknowledge the support of NVIDIA Corporation who provided access to GPU resources for the DFUC2020 and DFUC2021 Challenges.

\addtocmark[2]{Author Index} 
\renewcommand{\indexname}{Author Index}
\printindex

\bibliographystyle{unsrt}
\bibliography{Ref}

\end{document}